\documentclass[aip,jcp,reprint]{revtex4-1}

\usepackage[version=3]{mhchem} 
\usepackage{times}
\usepackage{graphicx}
\DeclareGraphicsExtensions{.pdf}

\begin{document}

\author{P. Wzietek}
\affiliation{Laboratoire de Physique des Solides, Universit\'e Paris-Sud 11, CNRS UMR 8502, 91405 Orsay, France}
\title
  {NMR spin-lattice relaxation in molecular rotor systems}

\date{\today}

\begin{abstract}
A general expression is derived for the dipolar NMR spin-lattice relaxation rate $1/T_1$ of a system exhibiting Brownian dynamics in a discrete and finite configuration space. It is shown that this approach can be particularly useful to model the proton relaxation rate in molecular rotors.
\end{abstract}

\maketitle 

\section{Introduction}

 Artificial molecular rotors are attracting much interest as a possible route to achieve a controlled motion at the nanoscale  \cite{Kottas2005,Kay07,Michl09,Vogelsberg12}.
A lot of progress has recently been made in solid-phase based systems as  molecular engineering permits to conceive crystalline environments where such rotors can move more-less freely 
\cite{Khuong06,Karlen10,Lemouchi11,Vogelsberg12jpc,Akutagawaa08,Lemouchi12,Rodriguez13,Jiang14,Lemouchi14}. 
 For molecular-sized rotors, even though a preferred direction of motion can be induced by application of an external driving field,   the main component of motion generally consists of random reorientation due to thermal vibrations of the crystal
lattice.

The experimental studies of such systems need tools sensitive to the dynamics of motion as complementary to average structure determination by X-ray diffraction. The latter 
gives mean positions of the atoms and the amplitudes of
their average displacement from the mean position,  however it does not give  information about the timescale of the displacement. Though, from the experimental point of view,  it is the study of the dynamics of the thermal motion which provides the main source of information about the interaction between a rotor and its environment. 
Among various experimental techniques that have been used for this purpose \cite{Kottas2005,Michl09,Vogelsberg12} NMR is probably the most powerful one as it can probe the dynamics with atomic resolution.  In crystalline-environment solid-state NMR techniques can be applied  similar to
standard liquid-state NMR methods that are commonly used to study dynamical processes in solution.

For crystalline systems most often basic NMR techniques are used, studying either the spectral shape or $T_1$ relaxation.
These two approaches differ by the range of frequencies of the motion that are probed.  The first relies on line-narrowing effects appearing when the frequency of the motion is comparable to the inhomogeneous NMR linewidth resulting from the local field anisotropy,
whereas the spin-lattice relaxation is sensitive to motion at the timescale of the Larmor frequency.  For example, deuterium NMR provides a very good probe as the lineshape is affected by strongly anisotropic quadrupolar interaction and simulation of 
$^2$H  NMR spectra has been often used in deuterated rotors \cite{Staehle14,Torres14,Vogelsberg12}.
However this technique needs deuterating the sample  (proton linewidth in solids is dominated by the dipolar proton-proton  interaction  which is less affected by a constrained anisotropic motion and the lineshape is not so easy to simulate). For protonated samples $^1$H $T_1$ relaxation provides instead a simple tool to study the dynamics \cite{Lemouchi11,Lemouchi11DT,Lemouchi12,Lemouchi13,Lemouchi14}.  

Proton spin-lattice relaxation by thermally activated motion, where the relaxation process is induced by a random modulation of dipolar interaction, is usually modelled using the well known Kubo-Tomita (KT) theory \cite{KT}. However the standard KT formula was derived for the simplest case of isotropic reorientation with dynamics described by a single correlation time, and for more complex systems the problem is to know the distribution of the correlation times and their relative weights in the $T_1$ relaxation rate. Such models were developed for polymers and biological materials \cite{Fatkullin,Korb}. 
This paper deals with relatively small systems that can be characterized by a finite number of correlation times, such as molecular rotors.

In the absence of an external driving field the thermal motion is determined by the torsional potential.  The equilibrium positions correspond to the potential wells and the nature of the motion depends on the height of the barriers $\Delta$ between neighboring wells with respect to the thermal energy $kT$ \cite{Kottas2005}.   
If the latter dominates the kinetics  of the motion is a Brownian random walk that can be modeled as a diffusion process (i.e. the probability of finding the rotor at a given position in function of time is governed by diffusion equation) where the diffusion constant depends on the effective friction and the inertia of the rotor. 
The model presented in this work present deals with the opposite case $\Delta>>kT$, probably more often encountered in crystalline environments. Here the motion is hindered by the barrier, in this condition the motion mainly consists of thermal librations within a well of the
torsional potential but the rotor also occasionally undergoes thermally activated hops between adjacent wells.
The thermally activated hopping rate $\Gamma$ is usually modeled by the Arhenius law 
\begin{equation}                       
\Gamma=\omega_0\exp -\Delta/kT 	
\label{eq:gamma}	
\end{equation}
where the attempt frequency $\omega_0$ is close to the librational frequency \cite{Kottas2005,Landauer61}.
The condition $\Delta>>kT$ implies $\Gamma<<\omega_0$ therefore, for small rotors in crystalline environment the spectral distribution of the characteristic frequencies of the motion is concentrated in two distinct regions.  This is in contrast to disordered polymeric materials or
biological macromolecules where the motion exhibits a rather continuous frequency spectrum \cite{Fatkullin,Korb}.
  Here the libration frequencies are typically of order of $10^{13}$Hz which is several orders of magnitude higher than the typical NMR frequencies therefore NMR is rather insensitive to this spectral component.
   On the other hand, the hopping rate $\Gamma$ can match the nuclear frequency at sufficiently low temperature leading to an efficient $T_1$ relaxation process. 

The fact that NMR is only sensitive to the slow spectral components of the motion is an interesting feature since it means that, as far as NMR is concerned, only the positions of the potential minima and the height of the barriers matter but the exact form of the potential is not relevant. This permits to make very simple and robust models for NMR $T_1$ relaxation. The first such model  has been proposed by Bloembergen et al. \cite{BPP} assuming that the random fluctuations of the local field due to hopping is described by a simple correlation function of the form
 $\exp -\Gamma t$. The  Kubo-Tomita  formula \cite{KT} results from the application of this model to the  case of relaxation by random modulation of the homonuclear dipolar coupling (see next section).  

The  simple exponential form of the correlation function is justified for hopping through a single potential barrier between two equilibrium states.  This theory applies well for e.g. hopping between two conformational states of a molecule.
 The purpose of this paper is to complete this theory with correlation functions calculated for an arbitrary number of degrees of freedom, wells and barriers. Examples that will be studied are rotors having different non-equivalent positions and a pair of coupled rotors.  
Inducing and detecting correlated motion in systems of coupled rotors is also one of the current challenges in the field, 
the studied example  may provide the background for understanding the results of a recent work where it was found that the $T_1$ relaxation  in a pair of coupled rotors is dominated by two different activated processes \cite{Lemouchi13,Bastien14}. 

In the literature devoted to molecular rotors, the term ''Brownian motion'' is often reserved to the case of purely thermal and unhindered
motion where the thermal energy $kT$ is dominant, that is the case $\Delta<<kT$.
But of course even in the opposite case considered here the hopping through the potential barriers is also a Brownian process.
In this regime, once we forget the librational degrees of freedom almost invisible to NMR, we are left with a system 
contained in a discrete configuration space (that is where space coordinates only take discrete values)   defined by the set of wells, exhibiting a Brownian dynamics.

The theory of the Brownian diffusion for rotating objects and the calculation of the correlation functions involved in NMR relaxation can be found in a number of works. For example, the textbook case of fully isotropic rotation (diffusion on a sphere) is treated in Abragam \cite{Abragam}, and a more general case of anisotropic rotation has been considered in  \cite{Favro60,Woessner62,Shimizu62,Shimizu64,Huntress68}. 
The calculation presented here follows the same philosophy except that it uses somewhat different mathematical tools adapted to the finite discrete space 
 and will therefore be detailed for completeness. 
Nevertheless it should be stressed that the theory presented here is in some sense more general because it is not merely restricted to a simple geometry of motion such as rotation.
The discrete configuration space  has no dimensionality in the usual sense (its topology is entirely defined by the set of hopping rates between different states), therefore the same formalism can be applied for {\em any} system that has a finite number of states, such as a pair of coupled rotors. 

The paper is organized as follows. In the next section we derive an expression relying the correlation functions to the hopping probabilities and apply it to calculate the dipolar relaxation rate $1/T_1$. In section \ref{sec:toy} we present two situations where this approach permits to go beyond the Kubo-Tomita analysis and calculate the variation of the relaxation rate $1/T_1$ from microscopic parameters.


\section{Theory}
\subsection{Basic formalism for motion-induced T1}
\label{sec:T1}

In this section we recall the basic formulas for the relaxation rate of a pair of nuclei induced by random modulation of their mutual dipolar interaction. After Abragam\cite{Abragam}, the general form of interaction hamiltonian can be written as
\begin{equation} 
{\cal H}_{int}(t)=\sum_q F^{(q)}(t)A^{(q)}
\label{eq:Hint}
\end{equation} 
where $F^{(q)}(t)=F^{(-q)*}(t)$ are some random functions of time related to the positions of the nuclei  and $A^{(q)}$ are spin variables. The index $q$ is usually related to the change in the $z$ component of the total spin for a given relaxation process. 
Note that  other mechanisms of relaxation induced by random motion, such as modulation of the quadrupolar interaction or the shielding tensor anisotropy, can be cast into this general form 
 \cite{Abragam,Shimizu62,Huntress68}. 
Here we focus on the relaxation by dipole—dipole interactions between identical spins, which is often the dominant source of relaxation in molecular rotor systems carrying protons (for the other cases we need to replace the Eqs.(\ref{eq:FsAs}) below by the appropriate expressions).

For this interaction, considering two identical spins $I_1$ and $I_2$ we have 
\begin{align} 
A^{(\pm1)}&=-\tfrac{3}{2}\gamma^2\hbar(I_{1z}I_{2\pm}+I_{1\pm}I_{2z})  \nonumber \\
  A^{(\pm2)}&=-\tfrac{3}{4}\gamma^2\hbar I_{1\pm}I_{2\pm} \nonumber \\
  F^{(\pm1)}&= \tfrac{1}{r^3} \sin\theta\cos\theta e^{\mp i\phi} \nonumber  \\
  F^{(\pm2)}&= \tfrac{1}{r^3} \sin^2\theta e^{\mp 2i\phi}  
  \label{eq:FsAs}
\end{align}  
Here the vector 
$\textbf{r}=(r,\theta,\phi)$ defines the relative positions of the spins and the time dependence of $F^{(q)}(t)$ comes from the random modulation of $\textbf{r}(t)$.  

Then the relaxation rate $1/T_1$ of the pair of spins  is related to the spectral density of fluctuations of $F^{(q)}(t)$ at the Larmor frequency $\omega_N$. Let us define the correlation function
\begin{equation} 
\Phi_q(t)=\left\langle F^{(q)}(\tau)F^{(q)*}(t+\tau) \right\rangle_{\tau}
\label{eq:def-Phiq}
\end{equation}
then the spectral density is
\begin{equation} 
J_q(\omega)=\int^{\infty}_{-\infty}\Phi_q(t)e^{-i\omega t}dt
\label{eq_Jq1}
\end{equation}
and the relaxation rate of the spin pair is given by
\begin{equation} 
1/T_1 = C(J_1(\omega_N)+J_2(2\omega_N))
\label{eq:T1vsJ}
\end{equation} 
with $C=\tfrac{3}{2}\gamma^2\hbar I(I+1)$.  
For an isotropic random rotation of the vector \textbf{r} characterized by a single correlation time $\tau$ the spectral densities are  \cite{Abragam,KT}
$J_1(\omega)\propto\frac{\tau}{1+\omega^2\tau^2}$ and $J_2(\omega)\propto\frac{4\tau}{1+\omega^2\tau^2}$
and the Eq.(\ref{eq:T1vsJ}) gives the well known Kubo-Tomita formula.

Note that the description of the system based on the hamiltonian (\ref{eq:Hint}) works under the assumption that the motion can be treated classically and that the orbital and spin variables are separated. This does not hold in the case where the system may undergoe quantum tunneling between configurations having different spin states. A well known example is the quantum tunneling observed in CH$_3$ groups, this process yields, at low temperatures, an additional relaxation channel  that is not captured within the hamiltonian (\ref{eq:Hint}) \cite{Allen69}.

\subsection{System dynamics: diffusion in a discrete space}
\label{sec:dynamics}

\begin{figure}[tb]
\begin{center}
\includegraphics[width=0.9\hsize]{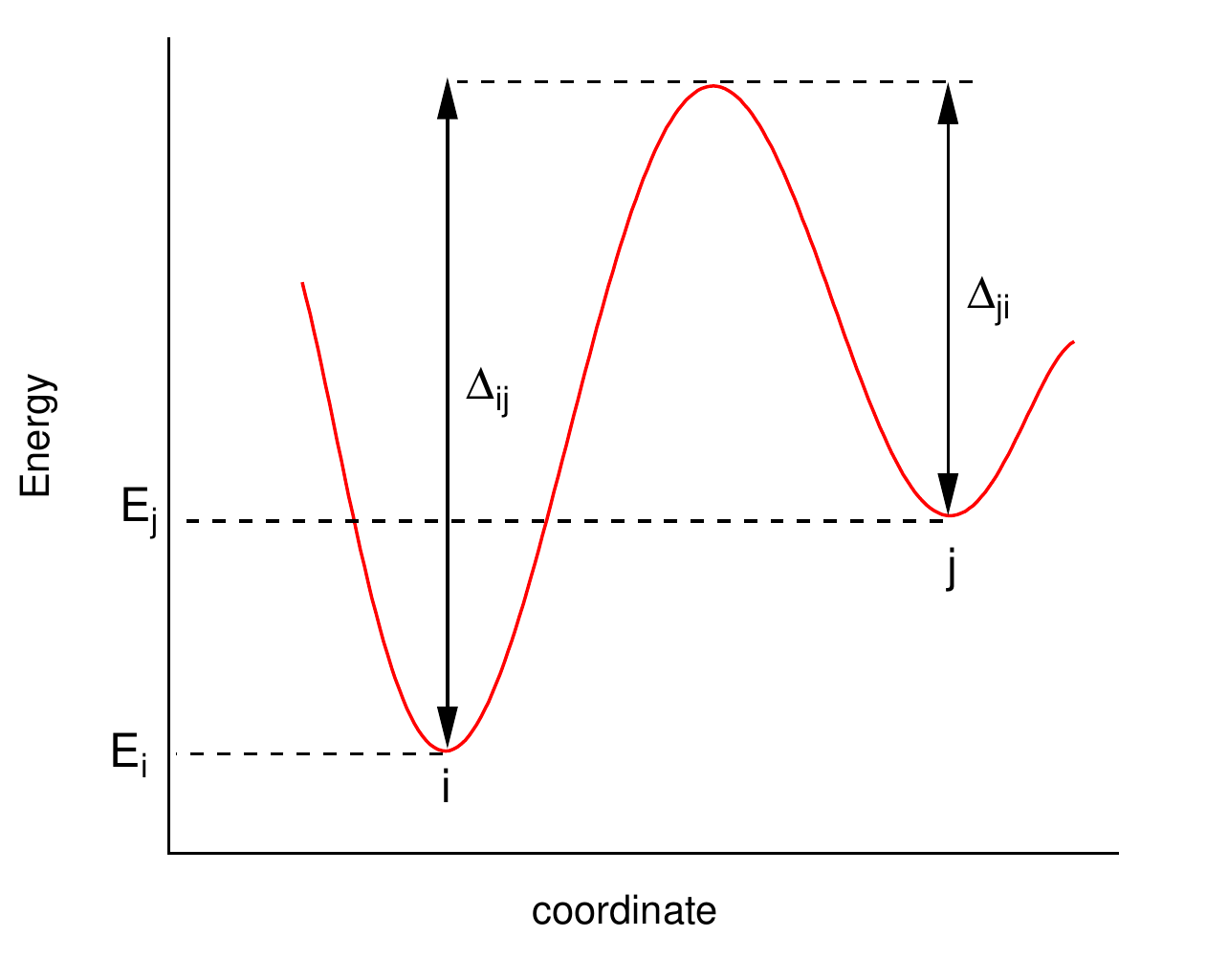}
\end{center}
\caption{The discrete model is defined by a set of potential wells $E_i$  and barriers $\Delta_{ij}$
between them. 
}
\label{fig:energy}
\end{figure}

At present we consider a system defined by a conformational potential exhibiting some number of wells of energy $E_i$ separated by barriers $\Delta_{ij}$, as sketched on Fig.\ref{fig:energy}.  The horizontal axis in Fig.\ref{fig:energy} represents schematically the degrees of freedom that may have more than one dimension (e.g. two angles for  a pair of rotors, as will be considered in the next section). As stated in the introduction we consider the case $\Delta_{ij}>>kT$, that is when NMR mainly senses the slow spectral component of the motion  due to thermal hopping between the wells.
Under such conditions the dynamics relevant for the $T_1$ relaxation can be modelled by diffusion in a discrete space defined by the set of states \{$E_i$\}. Let $n_i$ be the populations of these states,  it will be convenient to use the convention $\sum_in_i=1$ i.e. population is defined as the occupation probability. We denote as $n_i^{0}$ the populations at thermal equilibrium (then  $n_i^{0} \sim \exp(-E_i/kT)$)
and 
as $\Gamma_{ij}$ the probability of hopping $i\rightarrow j$ per unit time  (i.e. the average hopping rate), for a thermal process, according to (\ref{eq:gamma}) this probability is proportional to $\exp(-\Delta_{ij}/kT)$ where the barrier $\Delta_{ij}$ is defined in Fig.\ref{fig:energy}. Since $\Delta_{ij}-\Delta_{ji}=E_j-E_i$ we therefore also have  $\Gamma_{ij} = \Gamma_{ji} \exp(-(E_j-E_i)/kT)$. 

The time evolution of the system from a non-equilibrium state (relaxation) is governed by the master equation for populations:
\begin{equation}                       
\frac{dn_i}{dt} = \sum_{j\neq i} \Gamma_{ji}n_j(t)-\Gamma_{ij}n_i(t)		
\label{eq:master}	
\end{equation}
This equation can be rewritten as 
\begin{equation}
\frac{dn_i}{dt} = - \sum_j A_{ij}n_i(t)  
\label{eq:master_A}	
\end{equation}
with 
\begin{equation}
A_{ij}\equiv 
  \begin{cases}
   -\Gamma_{ji} & \ \ \textrm{for}\ i\neq j \\
    \sum_k \Gamma_{ik} & \ \ \textrm{for}\ i= j
  \end{cases}
\label{eq:def_A}	
\end{equation}
Then, in matrix notation with $\mathbf n=\{n_i\}$, $\mathbf A=\{A_{ij}\}$ the master equation is
$$
\frac{d\mathbf{n }}{dt}=-\mathbf{An}
$$
and has a general solution
\begin{equation}
\mathbf{n}(t)=e^{-\mathbf{A}t}\mathbf{n}(0)
\label{eq:ndetexpA}
\end{equation}
Therefore, the eigenvectors of $\mathbf A$ define the ''diffusion modes'' having different relaxation times related to the corresponding eigenvalues.

As an example, consider a rotor having $N$ equivalent positions (i.e. all states having the same energy and separated by the same barrier) and a uniform probability $\Gamma$ of hopping between adjacent states ($i\rightarrow i+1$ and $i\rightarrow i-1$).
Then the master equation reads
\begin{equation}
\frac{dn_i}{dt} = \Gamma\left(n_{i+1}+n_{i-1}-2n_i\right)
\label{eq:master_lin}	
\end{equation}
Note that the right hand side of this equation can be written as $\Gamma\left((n_{i+1}-n_i)-(n_i-n_{i-1})\right)$ showing that it is  obviously a discrete equivalent of the one-dimensional diffusion equation
$$
\frac{\partial n(x,t)}{\partial t}=D\frac{\partial^2 n(x,t)}{\partial x^2}
$$
For example, for $N=3$ (equilibrium positions at 0, 120 and 240 degree) the matrix $\mathbf A$  is
$$
\mathbf A =
\begin{pmatrix}
   -2 & 1 &1 \\
   1 &  -2 & 1\\
   1 & 1 & -2  
  \end{pmatrix}
$$
The corresponding eigenvectors  are:  $n_1+n_2+n_3$ with the eigenvalue 0 (the zero eigenvalue is always present because the total population number is conserved: $\sum n_i=$const), and  $n_1+n_2-2n_3$ plus permutations with the eigenvalue $3\Gamma$.  Thus $1/3\Gamma$ is the characteristic relaxation time of this system.

For a general $N$ it easy to show that the equation (\ref{eq:master_lin})	transforms to diagonal form with eigenvectors of the form $n_i\sim\cos ki$  yielding the eigenvalues $2\Gamma (1-\cos k)$
where the allowed values of the wave vector $k$ are determined by periodic boundary conditions 
$\cos(ki)=\cos(k(i+N))$, 
e.g. for $N=3$ these values are $k=0$, $2\pi/3$ and $4\pi/3$ corresponding to the eigenvalues $0$,$3\Gamma$,$3\Gamma$.

\paragraph*{Calculation of the correlation functions.}
Let $F$ be any function of state of the system, for a discrete configuration space $F$ is defined by the set \{$F_i$\}. Here we are interested in the dynamics of fluctuations of $F(t)$, defined in the equations (\ref{eq:FsAs}) (the index $q$ has been momentarily dropped to simplify general formulas), due to thermal hopping between different states of the system in  equilibrium.
This dynamics is contained in the autocorrelation function $\Phi$ of $F$ defined by the equation (\ref{eq:def-Phiq}). 
In a stationary random process  
$\left\langle F(\tau)F^*(t+\tau) \right\rangle=\left\langle F(\tau-t)F^*(\tau) \right\rangle$
which implies $\Phi(-t)=\Phi^*(t)$ therefore it will be convenient to restrict the analysis to the case $t>0$ and write the equation (\ref{eq_Jq1}) as
\begin{equation} 
J_q(\omega)=\int_{0}^{\infty}(\Phi_q(t)+c.c.)e^{-i\omega t}dt
\label{eq:Jq2}
\end{equation}

Let us denote as $P(i,t | i^\prime,t^\prime)$  the conditional probability that the system will be found in the state $i$ at time $t$ if we know that it was in a state $i^\prime$ at a previous time $t^\prime$.
For a stationary system this probability only depends on the time difference $t-t^\prime$, we therefore define a function $G_{ii^\prime}(t)$ as the conditional probability that the system is in state $i$ at time $t$ if it was in $i^\prime$ at $t=0$:
$$
G_{ii^\prime}(t)\equiv P(i,0 | i^\prime,t) 
$$
The matrix $\mathbf G (t)=\{G_{ik}(t)\}$ is the discrete-space equivalent of the Green function for the diffusion equation. It describes both (a) the relaxation and (b) the thermal fluctuations at equilibrium.
We will use (a) to find the relation between the matrices $\mathbf G (t)$   and $\mathbf A$,  
and then (b) to calculate the correlation functions.

From the definition of $\mathbf G (t)$ the relaxation of the system from an initial state described by populations $n_i(0)$ at time $t=0$ towards equilibrium is
$$
n_i(t) = \sum_j G_{ij}(t)n_j(0) 
$$
therefore, comparing this with Eq.(\ref{eq:ndetexpA}) one immediately sees that
\begin{equation}
\mathbf{G}(t)=e^{-\mathbf{A}t}
\label{eq:GvsA}
\end{equation}

The correlation function defined by Eq.(\ref{eq:def-Phiq}) describes the thermal fluctuations at equilibrium.
For stationary ergodic systems the time averaging can there be replaced by ensemble averaging at some fixed time e.g. $\tau=0$. In other words we can write:
\begin{equation}
\Phi(t)=\sum_{i,j}F_iF_j^* \  P(i,0; j,t) 
\label{eq:corrFens}
\end{equation}
where $P(i,0; j,t)$ stands for the probability of finding the system in the state $i$ at time $t=0$
\emph{and} in the state $j$ at some later time $t$.
Using $P(A;B)=P(B)P(A|B)$ this probability is $n_i(t=0)G_{ij}(t)$.  Here $n_i(t=0)$ is simply the equilibrium population $n_i^0$. Therefore the correlation function can be expressed in function of 
$\mathbf G (t)$ as
\begin{equation}
\Phi(t)=\sum_{i,j}F_iF_j^* n_i^0 G_{ij}(t)
\label{eq:corrFvsG}
\end{equation}

The expression for $1/T_1$ involves the Fourier transform of $\Phi(t)$ (Eq.\ref{eq:Jq2}).  This Fourier transform can be obtained by developping $\mathbf G (t)$ in the basis of the eigenvectors of $\mathbf A$.
Let \{$\mathbf M_m$\} be such a complete set  of eigenvectors and $M_{im}$ a matrix having these eigenvectors as columns, with the corresponding set of eigenvalues \{$\gamma_m$\}. The inverse eigenvalues $\tau_m\equiv 1/\gamma_m$ are the characteristic correlation times of the system
\footnote{Note that the eigenvalues are necessarily real even though $\mathbf A$ may not be symmetric.
}.

Then the equation (\ref{eq:GvsA}) can be written
\begin{equation}
G_{ij}(t)=\sum_{m}M_{im}M^{-1}_{\ \ \ mj}e^{-t/\tau_m}
\label{eq:GvsM}
\end{equation}
 and thus the correlation function (\ref{eq:corrFvsG}) becomes
\begin{equation}
\Phi(t)=\sum_{i,j,m}F_iF_j^* n_i^0 M_{im}M^{-1}_{mj}e^{-t/\tau_m}
\label{eq:PhivsM}
\end{equation}
Calculating the Fourier transform (\ref{eq:Jq2}) with  (\ref{eq:PhivsM}) we obtain finally
\begin{align} 
J_1(\omega)&=\sum_{i,j,m}n_i^0 M_{im}M^{-1}_{\ \ \ mj} (F^{(1)}_i F^{(1)*}_j+c.c.)\frac{\tau_m}{1+\omega^2\tau_m^2} \nonumber \\
J_2(\omega)&=\sum_{i,j,m}n_i^0 M_{im}M^{-1}_{\ \ \ mj} (F^{(2)}_i F^{(2)*}_j+c.c.)\frac{\tau_m}{1+4\omega^2\tau_m^2} 
\label{eq:Jqfinal}
\end{align}

The nuclear relaxation rate $1/T_1$ is given by the Eqs.(\ref{eq:T1vsJ}) with (\ref{eq:Jqfinal}) (where for practical calculations the zero eigenvalue i.e. $\tau_m=\infty$ should be excluded from the sum).
Note that the contributions of different states $i$  are weighted by their thermal occupation numbers $n_i^0$. This means that high energy states which are not often visited do not contribute significantly.

From Eq.(\ref{eq:Jqfinal}) it may seem that the number of distinct contributions that could a priori be distinguished in $1/T_1$ \emph{vs} temperature  is related to the the number of different correlation times $\tau_m$. However, since this expression involves sums $\sum_m M_{im}F^{(q)}_i$, the largest contribution will come from
 the eigenvectors $\mathbf M_m$ having the same symmetry as the sets $\{F^{(q)}_i\}$. 
The functions $F^{(q)}$ in (\ref{eq:FsAs}) are related to spherical harmonics of order 2 therefore we may expect that high-order diffusion modes do not contribute much to  $1/T_1$.  For example, for the model of Eq.(\ref{eq:master_lin})  it is found  that all but one eigenvector are orthogonal to $\{F^{(q)}_i\}$ so that the spectral densities (\ref{eq:Jqfinal}) are determined by a single correlation time.
This is somewhat similar to the situation found for the case of an isotropic rotation (diffusion on a sphere )\cite{Abragam} : in the expansion of the Green function in spherical harmonics only the terms of order 2 contribute to $1/T_1$ which is then characterized by a single correlation time.  This discussion explains thus the relative robustness of the standard Kubo-Tomita analysis even when applied to systems having several degrees of freedom. Examples where such simplification does not occur will be given in the next section.

The equations (\ref{eq:T1vsJ}) and(\ref{eq:Jqfinal}) have been derived for a single pair of spins. For abundant spins such as protons in organic materials where many spins can be coupled, the approximation considering only pairwise interactions is often justified by the fact that $1/T_1$ decreases as the $6^{th}$ power of the distance between spins so that it is dominated by the spin pairs that are the closest. On the other hand, long range dipolar coupling is often enough to ensure a common spin temperature and a single-exponential relaxation. This is for example the case of several compounds incorporating bicyclo[2.2.2]octane rotors where the closest proton pairs that dominate the relaxation are those of the CH$_2$ groups forming the rotor blades
\cite{Lemouchi11,Lemouchi12,Lemouchi13,Bastien14}. 
 In such conditions we can define an average relaxation rate i.e. 
the Eq.(\ref{eq:T1vsJ}) summed over proton pairs in a molecule. Likewise, summing the products $F^{(q)}_i F^{(q)*}_j$ in Eq.(\ref{eq:Jqfinal}) over the closest proton pairs of a rotor yields the contribution of a single rotor to the average relaxation rate.

\section{Toy models}
\label{sec:toy}
 
In this section we present two relatively simple situations where at least two distinct correlation times  
 are contributing to $1/T_1$. In both cases, although the variation of $1/T_1$ with temperature could also be modeled by a sum of two KT fits with arbitrary parameters,  our approach goes beyond the KT analysis because it tells exactly how the model parameters are related to the microscopic parameters such as the heights of the barriers and the positions of the wells
\cite{*[{All simulations have been done using IgorPro (www.wavemetrics.com), the fits of Fig.\ref{fig:T1-dbco}
were obtained using the Genetic Algorithm package, see }] [{.}] Nelson}.

\subsection{Model 1: single rotor with two barriers} 

\begin{figure}[tb]
\begin{center}
\includegraphics[width=0.9\hsize]{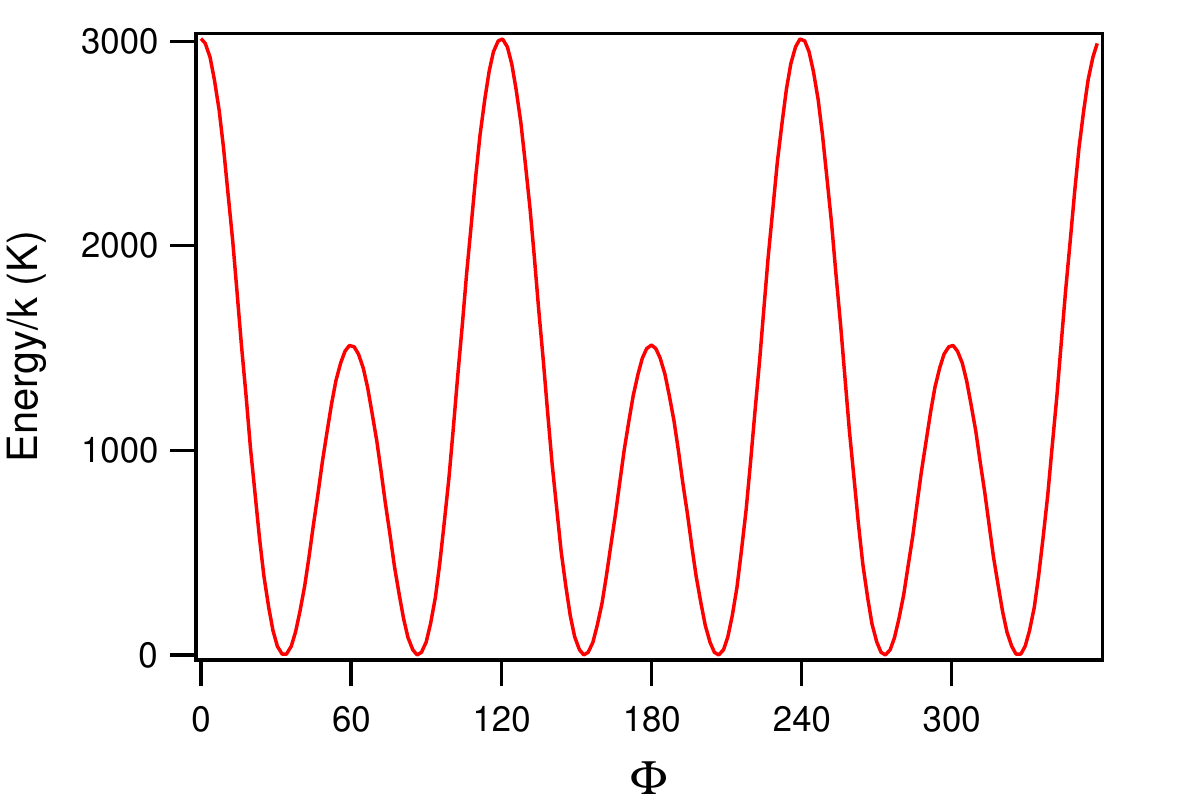}
\includegraphics[width=0.9\hsize]{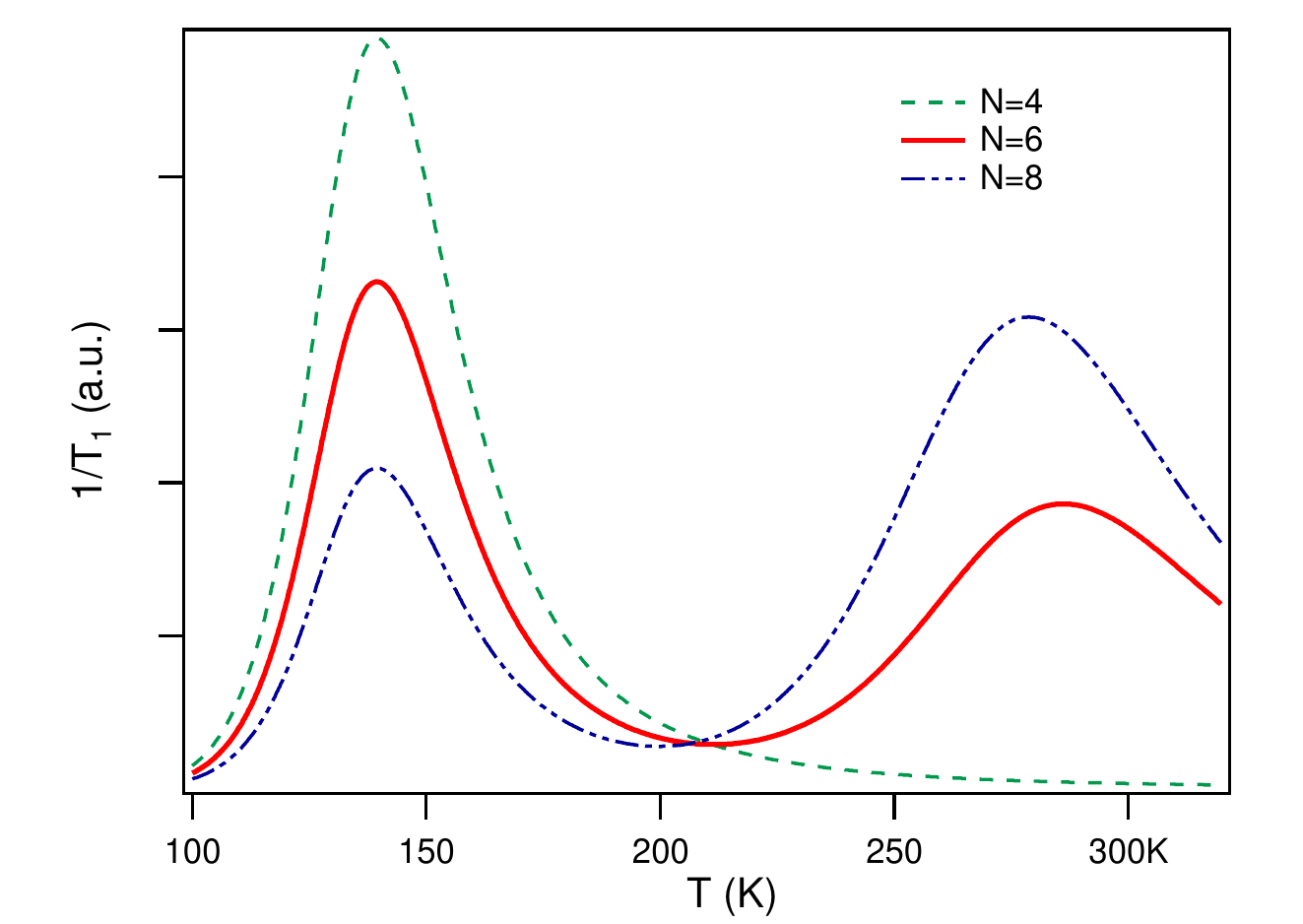}
\end{center}
\caption{{\em top}: Potential for model 1 with $N=6$. {\em bottom}: Temperature variation of $1/T_1$ for 
$N=4,6,8$.
}
\label{fig:V-T1-model3}
\end{figure}

We consider a single rotor with $N$ wells of the same energy separated by alternating small and big barriers such as shown in Fig.\ref{fig:V-T1-model3} (top) for $N=6$. 
For numerical calculation we have taken a sinusoïdal potential with arbitrarily chosen values of 1500~K and 3000~K for the barriers (since we are interested in the temperature variation it is natural to express all energies in Kelvin, 1kcal/mol$\sim 503$~K), $10^{13}$s$^{-1}$  for the attempt frequency  and 50~MHz for the NMR frequency.

Fig.\ref{fig:V-T1-model3} (bottom) shows the simulations for three values of $N=4,6,8$. The two peaks, around 140~K and 280~K,  correspond to the activation of motion through the small and big barrier respectively. Since the low temperature process is a restricted rotation the corresponding relative amplitude 
decreases with increasing N (e.g. for N=8 the $\sim45$ degree jumps are much less efficient in nuclear relaxation than the 90 degree jumps). The amplitude ratio is in general quite sensitive to the position of the potential minima which depend on the degree of anharmonic terms in the potential.

\subsection{Model 2: Two rotor system and correlated motion}

This model is inspired by recent works \cite{Lemouchi13,Bastien14} where the rotors are 
bicyclo[2.2.2]octane (BCO) functional units embedded in crystalline lattice \cite{Lemouchi11,Lemouchi12}.
The nuclear relaxation generated by the rotating  BCO rotors comes from the modulation of the dipolar coupling between the proton pairs of the CH$_2$ groups forming the rotor blades. 
The interesting situation arises when pairs of such rotors are close enough such that the interaction 
between them adds a significant contribution to the potential barriers favoring a correlated motion.
A system with pairs of crystallographically equivalent, interacting BCO rotors has been synthetized by Lemouchi et al.\cite{Lemouchi13}.
Proton  relaxation rate study has shown the existence of two different activated processes of very different energies
and it was conjectured that these two processes could be attributed to  disrotatory and conrotatory modes of rotor pair rotation \cite{Lemouchi13,Bastien14}.

In the Appendix \ref{sec:app} we address the question whether
the existence of a correlated mode of motion can lead to appearance of a distinguished activated process seen in NMR relaxation.
 Intuitively this could be expected if we consider the relaxation by inter-rotor dipolar coupling, however it is less obvious if we only consider proton pairs sitting on the same rotor. 
As far as the considered compound is concerned    the intramolecular H-H distances are always the shortest
\cite{Lemouchi13} therefore we only consider the intra-rotor spin pairs
\cite{*[{Note that for inter-rotor dipolar coupling the approximation of independent proton pairs does not hold anymore, for example due to the threefold symmetry each proton exhibits the same interaction with at least three other protons on the neighboring rotor.  Such mechanism  has been revealed and studied in hydrated
gypsum, where relaxation is due to interaction between protons of neighboring water molecules, see }] [{.}] Holcomb, *Jeener1,*Jeener2}.

\begin{figure}[tb]
\begin{center}
\includegraphics[width=0.4\hsize]{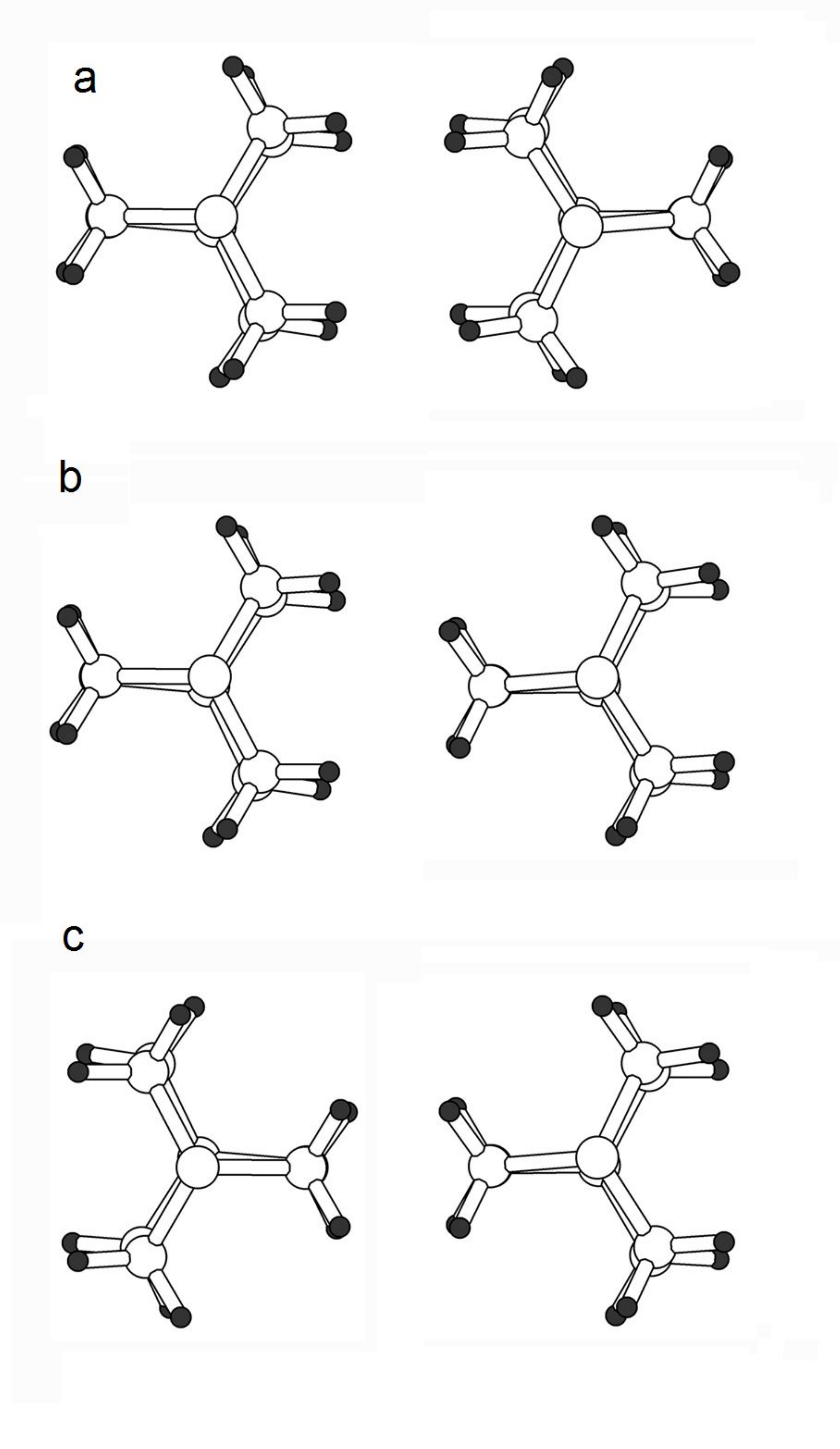}
\includegraphics[width=0.9\hsize]{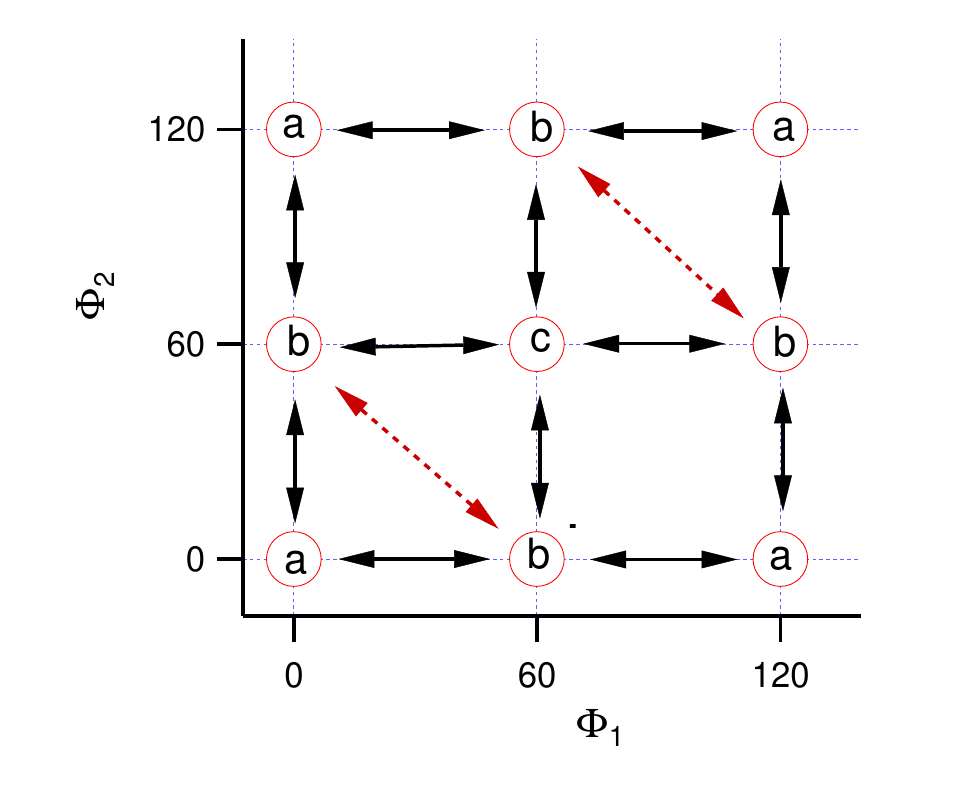}
\end{center}
\caption{{\em top}: Possible  configurations $(a,b,c)$ of a pair of interacting BCO rotors, after Lemouchi et al. \cite{Lemouchi13}. {\em bottom}: The conformational map of the two-rotor system is defined by two angles ($\phi_1$, $\phi_2$). The map is periodic by 120 degree  due to the threefold symmetry of the rotors.
In the discrete model used here we consider the equilibrium states  $a$, $b$  and $c$ and transitions between them: 
full arrows represent the single rotor jumps and dashed arrows the correlated jumps.  
}
\label{fig:V-model2D}
\end{figure}

 For a pair of interacting rotors we consider the set of the three
  equilibrium configurations as depicted in Fig.\ref{fig:V-model2D} (top),  with majority-majority ($a$), majority-minority ($b$), and minoroty-minority (Fig. $c$) 
occupations  of the rotor positions in a pair determined in the room-temperature crystal structure as discussed by Lemouchi \emph{et al.} \cite{Lemouchi13}. 
 The resulting conformational map is  shown in Fig.\ref{fig:V-model2D} (bottom) where the arrows correspond to the transitions considered in our model. The  correlated jumps are those between neighboring $b$ states where both rotors exhibit a gear-like rotation by 60 degrees.
Here we are studying the simplest model where the potential barriers are only due to rotor-rotor interaction, this model is parametrized by three energy barriers $\Delta_{ab}$, $\Delta_{bb}$, $\Delta_{bc}$
(the map would have lower symmetry in the general case where the crystalline environment of each rotor is taken into account).

\begin{figure}[htb]
\begin{center}
\includegraphics[width=0.95\hsize]{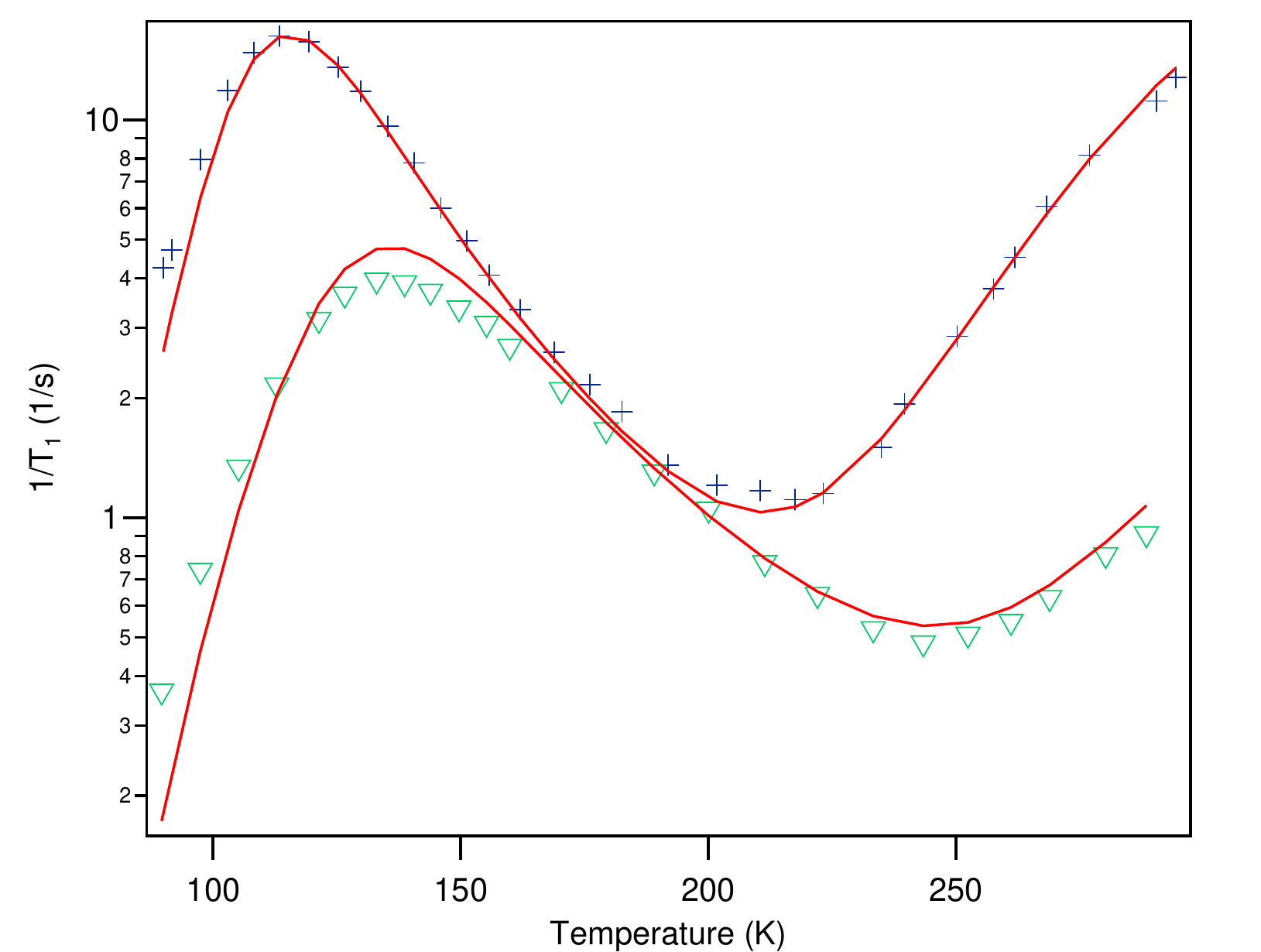}
\end{center}
\caption{Symbols: $1/T_1$ data at 50MHz (crosses) and 210MHz (triangles) from Lemouchi et al.\cite{Lemouchi13}.
Solid lines: simulations using the model of Fig.\ref{fig:V-model2D} (see text).  
}
\label{fig:T1-dbco}
\end{figure}

In Fig.\ref{fig:T1-dbco} we show the results of simulations.  The solid line shows the best fit with the experimental data of  Lemouchi et al. \cite{Lemouchi13} obtained with 
$E_b-E_a=140$K (0.28kcal/mol), $E_c-E_a=1180$K (2.35kcal/mol), 
$\Delta_{ab}=3170$K ($\sim$6.3 kcal/mol),    
$\Delta_{bb}=1000$K($\sim$1 kcal/mol), %
$\Delta_{bc}=1500$K($\sim$3 kcal/mol) and the attempt frequencies of 1.9~$10^{12}$s$^{-1}$ and 8.2~$10^{11}$s$^{-1}$ for the uncorrelated and correlated jumps respectively. We could not find any significantly different set of parameters reproducing this data.
Within this set the high temperature activated process would be related to the high value of $\Delta_{ab}$, that is of the process of breaking the majority configurations $a$.
This assignment seems to be supported by Carr-Parinello simulations\cite{note-CPsimul} showing a relatively long  residence time in the $a$ configurations compared to that of other states. 
As for the low temperature process characterized by $\Delta_{bb}$,  
it requires quite a low value of the energy of $b$ configuration ($E_b-E_a$) to account for the observed $1/T1$ data. Such a low value is however not consistent with the estimation of the occupation rate of $b$ sites  inferred from X-ray studies \cite{Lemouchi13}. At any rate this simple model, where only rotor-rotor interactions play a role and the crystalline potential is neglected, is probably oversimplified so that the  simulations shown in Fig.\ref{fig:T1-dbco} should be  considered  as toy model simulations rather than a fit to the data.

\section{Conclusions}

In this work we have proposed a simple and general approach to calculate 
correlation functions involved in nuclear spin-lattice relaxation for a system undergoing thermal hopping between a number of potential wells. We argue that there should be many cases where this approach can yield a much more complete information about the dynamics of motion compared to that obtained from a simple Kubo-Tomita fit. 
Examples of  this approach in the field of molecular rotors are studied where the analysis of the temperature dependence of the relaxation rate may provide detailed information about the form of the conformational  potential. In particular, the method is capable of modelling the effects of gear-like motion in coupled rotor systems. The use of this approach might be generalized if numerical tools are developed similar to packages widely used to study dynamics by simulation of NMR spectra.


\begin{acknowledgments}
I thank P.Batail for collaboration on related issues and for giving me inspiration for this work.
I also thank E. Canadell, C.Lemouchi and  P.Kalugin   for enlightening discussions.

\end{acknowledgments}

\appendix
\section{Relaxation by intramolecular spin pairs and correlated rotors}
\label{sec:app}

 The $T_1$ due to the intramolecular spin-spin interaction obviously probes only the {\em single rotor correlation function} it is therefore interesting to ask whether it can reveal the existence of a correlated mode of rotation.
It will be convenient to define a composite index $(i,i\prime)$ where $i$ and $i\prime$ number the position the left and right rotor respectively.
Then, considering the relaxation of a proton pair on the left rotor, the involved geometrical factors are independent of $i\prime$:
$F^{\textrm{left}}_i\equiv F_{(i,i')}$
which implies that 
the correlation function will involve the effective $\mathbf G$ matrix for the left rotor, defined as a contraction of $\mathbf G$ over the indices of the right rotor:  
$$
G^{\textrm{left}}_{ij}=\sum_{i',j'}G_{(i,i')(j,j')}
$$
Likewise we can write the master equation for the effective population of the left rotor:
$$
\frac{dn^{\textrm{left}}_i}{dt} =\frac{d}{dt}\sum_{i'}n_{(i,i')}(t) = - \sum_{i',j,j'} A_{(i,i')(j,j')}n_{(j,j')}(t) 
$$
Then, if we consider a model where the hopping probability of the left rotor does {\em not} depend on the current
position of the right rotor, that is if we only consider hopping between crystallographically equivalent states (e.g., in our case 120 degree jumps) then 
$\sum_{i'} A_{(i,i')(j,j')}$ does not depend on $j'$. In such case, whether correlated jumps exist or not,  the effective master equation describes a single rotor problem with an effective hopping rate:
$$
\frac{dn^{\textrm{left}}_i}{dt} = - \sum_{j}\left(\sum_{i',j'} A_{(i,i')(j,j')}\right)n^{\textrm{left}}_j(t) 
$$

This means that, even if a correlated mode of rotation exists, it does not necessarily lead to distinct activated process that could be seen in the temperature dependence of $1/T_1$. In our case, such  correlated mode of rotation between the neigboring $a$ states would just renormalize the single-rotor correlation time without introducing any new correlation times. 
%
%
Therefore, if the model is to account for a distinct and observable process related to the correlated motion we must take into account all non-equivalent configurations so that  the hopping probability of a rotor depends on the current
position of the neighbouring rotor, as in the set of states shown in Fig.\ref{fig:V-model2D}.

%


\bibliography{rotors}

\end{document}